\documentclass[aps,prb,twocolumn,groupedaddress,showpacs,amsfonts]{revtex4}
\usepackage{graphicx}
\begin{document}

%\preprint{}

\title{Structure, Photophysics and the Order-Disorder Transition to the Beta Phase in Poly(9,9-(di{\it -n,n}-octyl)fluorene)}
\author{M.J.~Winokur}
\email[]{mwinokur@wisc.edu}
\homepage[]{http://romano.physics.wisc.edu}
\author{J.~Slinker}
\author{D.L.~Huber}
%\thanks{}
\affiliation{Department of Physics, University of Wisconsin, Madison, WI 53706 }


\date{\today}

\begin{abstract}
X-ray diffraction, UV-vis absorption and photoluminescence (PL) spectroscopy have been used to study the well-known order-disorder transition (ODT) to the $ \beta $ phase in poly(9,9-(di{\it -n,n}-octyl)fluorene)) (PF8) thin film samples through combination of time-dependent and temperature-dependent measurements. The ODT is well described by a simple Avrami picture of one-dimensional nucleation and growth but crystallization, on cooling, proceeds only after molecular-level conformational relaxation to the so called $\beta$ phase.  Rapid thermal quenching is employed for PF8 studies of pure $\alpha$ phase samples while extended low-temperature annealing is used for improved $\beta$ phase formation.  Low temperature PL studies reveal sharp Franck-Condon type emission bands and, in the $\beta$ phase, two distinguishable vibronic sub-bands with energies of approximately 199 and 158 meV at 25 K.  This improved molecular level structural order leads to a more complete analysis of the higher-order vibronic bands.  A net Huang-Rhys coupling parameter of just under 0.7 is typically observed but the relative contributions by the two distinguishable vibronic sub-bands exhibit an anomalous temperature dependence. The PL studies also identify strongly correlated behavior between the relative $\beta$ phase 0-0 PL peak position and peak width.  This relationship is modeled under the assumption that emission represents excitons in thermodynamic equilibrium from states at the bottom of a quasi-one-dimensional exciton band.  The crystalline phase, as observed in annealed  thin-film samples, has scattering peaks which are incompatible with a simple hexagonal packing of the PF8 chains.
\end{abstract}

\pacs{61.30.Vx,61.10.-i, 78.66.Qn, 81.30.Hd, 82.35.Cd}
%\keywords{}

\maketitle


% If in two-column mode, this environment will change to single-column
% format so that long equations can be displayed. Use
%\begin{widetext}
%\end{widetext}
\section{Introduction}

Conjugated oligomers and polymers continue to attract world-wide attention because of the great promise of low-cost, small-format optical and electronic device applications\cite{CP:applications,dev:theory:advmat01bredas}. This goal is heavily reliant on new materials based, in part, on the synthetic addition of solubilizing side chains and the facile processing technology that follows\cite{PAT:McCullough:98reviewAM,dev:inkjet:sci00Friend}.  These chemical modifications have also created a vast wealth of new structure/property relationships which modify important physical properties including both photophysics and charge transport and, as a consequence, those of fabricated devices.  Much current research is aimed simply at understanding the impact that even simple post-synthesis processing treatments have on observed polymer properties.  
  
Especially interesting are conjugated polymers incorporating poly({\it p}-phenylene) based backbones.  These materials possess relatively large band gaps and are considered suitable for applications requiring blue emission\cite{PFO:blueLED:apl98Bradley,PFO:CIEblue:aplscherf01,PFO:bluereview:macrorc01neher,PFO:HEffLED:aplCacialli01}. Of these, polyfluorene derivatives (PFO's) are known\cite{PFO:twenty:jps_Leclerc01,PFO:LEDreview:macroRC_Neher01,PFO:thermal:mmZeng02} for having excellent quantum efficiencies, good electron mobilities and exceptional thermal and chemical stability in inert environments.  Recent work\cite{PFO:photo_lowrelax:sm_vardeny01} has focused on the properties of a small number of linear and branched dialkyl-substituted fluorenes.  Poly(9,9-(di{\it ~n,n}-octyl)fluorene), or PF8 as schematically shown in Fig.~1, has received extra attention because it exhibits mesomorphic behavior and multiple crystal phases\cite{PFO:PLpolar:prb99bradley,PFO:alignlcp:apl99Bradley,PFO:processXRD:mm99Bradley,PFO:F8morph:prb00Bradley,PFO:NSOM:langTeetsov02}. On heating of PF8 thin films one generally observes transitions from a crystalline state to a liquid crystalline mesophase followed by melting to an isotropic phase.  Consequently there are special opportunities\cite{PFO:laserprocess:apl99Vardeny,PFO:Ramanmorph:sm00Bradley,PFO:alphabeta:cpl_Vardeny02,PFO:F6F8raman:jap02bradley} for studying the response of electronic and optical properties to systematic changes in the molecular level ordering.  Mesomorphic behavior is seen in many other conjugated polymer families containing alkyl side chain  substituents\cite{psil;t-chrom1,PAT;chrom;shu87,PAT;mjw;89} although the exact details are both backbone and side chain specific.  It also extends to oligomers and oligofluorenes bearing chiral side chain substituents\cite{PFO:oligo:jpc_leclerc02,PFO:chirallcp:mmScherf02,PFO:circpolpolymorph:jacs02chen} and these show very rich phase behavior.

\begin{figure}
\includegraphics[width=3.2in]{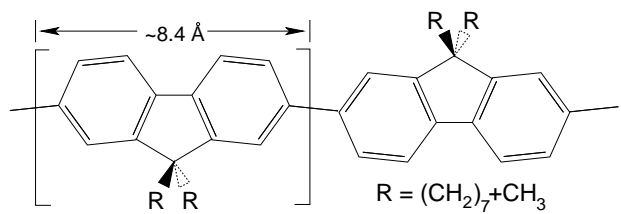}
\caption{A sketch of the poly(9,9-(di{\it ~n,n}-octyl)fluorene) or the PF8 polymer.\label{fig1}}
\end{figure}

The mesomorphic behavior of the PF8 polymer correlates with the existence of two distinct backbone conformations\cite{PFO:processXRD:mm99Bradley,PFO:abphase:actapolym98Bradley} referred to as the ``alpha'' or ``beta'' phase\cite{conformer}.  These two forms are distinguished by an approximate 0.1 eV difference\cite{PFO:beta100mev:sm00Bradley} in the interband $\pi$~-~$\pi^*$ transition and this effect is seen in both optical absorption and emission spectra.  The shorter wavelength $\alpha$ form is claimed to have increased conformational disorder, a less planar backbone structure and, with respect to the interchain packing, reflective of a glassy type state.  The explicit backbone conformations are still unknown.  A branched side chain polyfluorene derivative, poly(9,9-di(ethylhexyl)fluorene) or PF2/6, only exhibits optical properties relating to the $\alpha$ form.  Diffraction studies by Lieser et al.\cite{PFO:scherf:mm00helix52} are therefore somewhat surprising because this polymer clearly adopts a well-ordered structural  phase with good evidence for 5$_1$ or 5$_2$ fluorene helices.

At temperatures above 80~$^\circ$C the PF8 polymer undergoes a reversible order-disorder transition (ODT) to a polymer liquid crystal (PLC) phase consisting entirely of the $\alpha$ conformation\cite{PFO:processXRD:mm99Bradley}.  On cooling the polymer reverts back to the $\beta$ type conformer although a substantial fraction of the $\alpha$ form persists.  In bulk PF8 samples\cite{PFO:processXRD:mm99Bradley} this more ordered state has been associated with crystallization.  Thin films also manifest signatures of this ODT but they also exhibit X-ray scattering patterns indicative of a completely different unit cell structure\cite{PFO:tfXRD:polyBradley02}.  Formation of the $\beta$ type conformer is also observed in very dilute PF8/polystyrene blends\cite{PFO:processXRD:mm99Bradley} upon exposure to toluene vapor.  While the ODT clearly occurs at temperatures above the cited 75~$^\circ$C glass transition temperature there are reports of increased conversion to the $\beta$ form and improvements in the photophysical properties from samples cycled from 300 K down to 77 K and back to room temperature\cite{PFO:F8morph:prb00Bradley}.   

Beyond the unresolved questions concerning the fundamental relationship between the ODT, crystallization and formation of the so called $\beta$ phase there are a wealth of other important issues.  PFO's typically include significant heterogeneity and disorder and both optical and transport data clearly reflect consequences of these effects. Compared to short oligomers most polymer optical data include very broad spectral features, an effect termed inhomogeneous broadening, and this property further complicates interpretation of these data in many instances.  To circumvent this limitation, and to better understand the nature of the photophysics, both time-resolved and site-selective spectroscopy are used\cite{SSF:PPV:prb96friend,SSF:fsPPV:jcp97Leising,SSF:review:acr99Bassler,PPV:trSSF:prb01phillips}.  It has also proven useful to investigate isolated oligomers while frozen within various inert organic matrices\cite{SSF:review:acr99Bassler}.  Ultimately one is interested in discerning intrinsic properties of the polymers from those which are simply reflective of residual molecular level disorder.  

This communication addresses the nature of the ODT in the PF8 polymer and its relationship to formation of the $\beta$ phase and crystallization.  Our most important result in this respect is to demonstrate that formation of the $\beta$ phase has properties consistent with a simple picture of nucleation and one-dimensional growth but that overt crystallization occurs as a secondary process and, depending on the temperature, with much longer characteristic time frames.  A second, seemingly fortuitous result demonstrates that much of the low-temperature spectral broadening in the PF8 polymer is an extrinsic property that relates only to the processing history.  Emission spectra, from cast-films having reduced disorder, exhibit a marked vibronic structure at low temperatures.  These enable a better overall analysis of the Franck-Condon type vibronic signatures and a clearer assessment of the overall impact of processing history in temperature-dependent studies.  

\section{Experimental Method}

Synthesis of the poly(9,9-(di{\it -n,n}-octyl)fluorene is reported elsewhere\cite{PFO:synth:chemrev95Suzuki,PFO:synth:SPIE99Bernius}.  The PF8 polymer used in this work was acquired from American Dye Source (polymer BE129) and used as received.  An unfiltered 1\% w/w solution of polymer in tetrahydrofuran (THF) was prepared and then placed in an ultrasonic bath for over one hour to maximize dissolution. Some reduction in the molecular weight may have occurred\cite{lessdisorder} but the absorption and emission maxima were comparable to previously published values. For optical studies the solution was initially drop-cast or spin-coated onto 100 $\mu$m thick sapphire substrates.  Because toluene is considered to be a good solvent the 1\% solution was subsequently diluted to ca.\ 0.5\% w/w by addition of toluene and again drop-cast or spin-coated (at 3000 rpm for 60 sec).  All polymer films were dried in air at ambient conditions and then mounted for further study.  Thick films appeared soft and ``gummy'' for extended periods after casting suggestive of significant residual solvent.  In this paper we include results from three different samples: A, a drop-cast film from the 1\% PF8 w/w THF solution; B, a drop-cast film from the 0.5\% w/w THF/toluene mixture; C, a spin-cast film from the 0.5\% w/w THF/toluene mixture.

Steady-state UV-vis absorption (Abs) and photoluminescence (PL) spectroscopies were performed using a custom-built spectrometer equipped with a vacuum oven/cryostat chamber.  This chamber also included a nozzle port which could direct a pressurized jet of CO$_2$ spray onto the sapphire substrate for {\it in situ} quenching experiments.  Cooling rates down to a base temperature of ca.\ -50~$^\circ$C were estimated to be well in excess of 25~$^\circ$C/sec.  Once quenched, samples could be further cooled using a cryostat cold finder (either liquid N$_2$ or, less frequently, a closed-cycle He displex).   A single array spectrometer (Ocean Optics USB2000, range 200-800 nm with better than 2.8 nm FWHM resolution) was used to sequentially acquire both absorption and PL spectra from superimposed illumination areas.  For excitation a 150 W Xe lamp (Oriel 4220) was coupled to a primary monochromator (JY HT 20) and then focused to give a 0.4 $\times$ 0.8 mm spot on the polymer film.  A focusing fiber optic cable was configured to pipe white light (from the Xe lamp) onto the sapphire substrate from behind.  The sample chamber was equipped with a manually operated X-Y-Z translation stage.  

All sapphire substrates included areas that were kept polymer free.  These locations were then used as a reference point for the absorption measurements.  Absorption variations of less than 0.002 could be reproducibly resolved. Spin-cast films required additional treatment to obtain a polymer free area.  This was achieved by dipping the substrate halfway into pure toluene multiple times.  The dipping was arranged so that there was always an intermediate strip with much reduced film thickness.  Typical PL or Abs spectra required five seconds or less of data acquisition time.  Because both absorption and PL were acquired the PL data could be corrected for self-absorption. Even in the worst cases (i.e., ``thick'' samples in the $\alpha$ phase) the overall effects of this correction at energies below 3.0 eV  were relatively minor.  Data that include self-absorption corrections are indicated in the respective figure captions.

X-ray data was recorded using a powder diffractometer, based on an Inel CPS-120 position sensitive detector, mounted to a 15 kW rotating anode X-ray generator (Cu K$_\alpha =1.542$ \AA). All beam paths were He gas-filled to minimize absorption and air scatter. For this work the goal was simply to reproduce as closely as possible the conditions used in the optical spectroscopy while compensating for the limited sensitivity of the X-ray diffractometer in transmission mode geometry.  Because of the weak scattering signal from the polymer it was necessary to increase the film thickness by drop casting between 10 and 20 times onto the same spot area.  To minimize absorption effects the polymer was cast onto $\approx 5 \mu$m thick mica supports instead of sapphire.  The final film was rather inhomogeneous and had an estimated thickness of 5 $\mu$m.  The sample was mounted onto a Peltier equipped sample stage (for temperatures ranging from -35 to 120~$^\circ$C).  Data acquisition times ranged from 2 to 8 hours per data set.

\section{Results and Discussion}

In Fig.~2 we show a typical series of PL emission data and, in the inset, the two limiting absorption spectra from a THF-only solution (sample A) drop-cast PF8 film recorded on cooling (ca.\ 3 K/min) from 350 K to 25 K.  The majority of these data are comparable to previously published reports.  At temperatures above 130 K the PL is dominated by three moderately broad emission features and, at 350 K, these are centered near 2.87, 2.68,and 2.52 eV.  They correspond to singlet exciton decay and, within a simple Franck-Condon (F-C) picture, are indicative of 0-0, 0-1 and 0-2 vibronic transitions between the lowest level in the 1$^{\mbox {st}}$ excited electronic state to various vibronic levels within the electronic ground state manifold.  There is a weak shoulder near 2.3 eV and it is likely dominated by the 0-3 vibronic band transitions.  In support of this we note that this shoulder exhibits temperature-dependent energy shifts which parallel those of the more intense vibronic peaks.  Emission at
chemical defects is also a possibility.  The presence of keto type defects is thought to be low in this sample because its characteristic emission\cite{PFO:PLdefect:apl02meijer,PFO:keto:advmatList02} is typically centered near 2.3 eV and there is little or no fixed (i.e., temperature independent) emission at this energy.  The F-C type features appear to be superimposed on an extremely broad emission ``band'' which is centered near 2.4 eV.  This underlying background originates from a variety of potential experimental sources including residual $\alpha$ phase emission, unresolved vibronic replicas and interchain (IC) excitations\cite{PFO:PLdefect:apl02meijer,privcomm:guha} in addition to the
chemical defects just discussed.  The background profile can also be model dependent\cite{better}.
\begin{figure}
\includegraphics[width=3.375in]{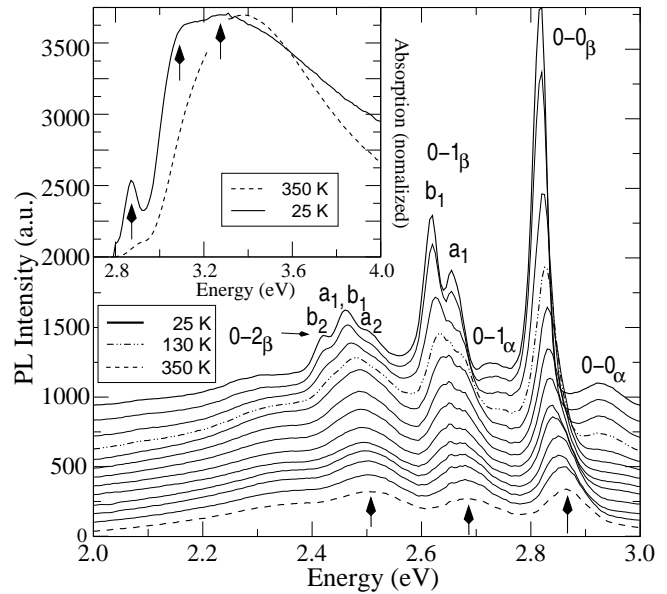}%
\caption{Progression of PF8 sample A photoluminescence spectra on cooling from 350K   ($\lambda_{\mbox{\tiny{Ex}}} = 390$~nm).  All PL curves are self-absorption corrected and offset for clarity.  Inset: Two corresponding absorption spectra with arrows indicating the $\beta$ phase 0-0, 0-1 and 0-2 vibronic bands.\label{fig2}}
\end{figure}

Intrachain exciton migration\cite{mLP:emigration:cplett00List} is always present so that the prompt PL\cite{PFO:prompt:jpc01Bassler} is strongly dominated by emission from fluorene segments with the longest effective conjugation lengths.  Interchain energy transfer processes can also be significant\cite{PFO:interchain:pnas02belijonne}.   Absorption, at sufficiently short incident wavelengths, occurs throughout the polymer and so better reflects the average conformational structure (assuming, of course, that all interband absorption cross-sections are nearly equal).  At 350 K there is only a weak shoulder at 2.95 eV, associated with the interband absorption from the $\beta$ conformer, and a much broader absorption centered at 3.38 eV characteristic of the $\alpha$ form.  On cooling there is a significant increase (not shown) in the $\beta$ phase absorption and, at 25 K, the two higher energy F-C vibronic bands, again a 0-1 and 0-2, are resolved.  In this case the transition is between the ground state and vibronic levels within the 1$^{\mbox {st}}$ excited electronic state manifold.  Although the PL is always dominated by $\beta$ type contributions the absorption spectra always indicate that a large proportion of fluorene backbones remain trapped in the high-temperature $\alpha$ form.

The low temperature PL is far more interesting. The most notable feature that develops is an excessive narrowing of the $\beta$ phase peaks.  After correcting for the finite 18 meV instrumental resolution, and then assuming a simple convolution of Gaussians\cite{better2}, we arrive at an estimated 18$\pm 1$ meV 0-0 overall line width for the PF8 polymer at 25 K.  This width is many times narrower than is typically seen in other PFO's and, as far as we are aware, is only rivaled in prior polyfluorene studies that have utilized site-selective fluorescence (SSF) spectroscopy\cite{PFO:timessl:cpl01Bassler}.  In addition there are also new, relatively weak peaks at 2.93 and 2.73 eV.  This paper will later show that these features are primarily due to emission from 0-0 and 0-1 bands in residual $\alpha$ type regions of the polymer. 

At this sharply reduced width both the 0-1 and 0-2 phonon bands manifest a pronounced sub-band structure with two and three distinguishable components respectively.  Splitting of the 0-1 band is resolved in SSF studies\cite{PFO:timessl:cpl01Bassler} as well.  Vestiges of the 0-3 sub-band structure also appear. The overall PL line shape of this polymer bears a striking resemblance to high resolution studies of small phenylene vinylene oligomers\cite{OPV:subband:jcp02Gierschner} (OPV's) and this aids in the identification of the various salient features. The more intense 0-1 sub-band, at 2.618($\pm 0.002$) eV, is identified with an in-plane deformation mode of the backbone and its $\sim$199 meV offset from the 0-0 band, at 2.817($\pm 0.002$) eV, matches the energy of a very strong and well isolated 1600 cm$^{-1}$ Raman band\cite{PFO:Ramanmorph:sm00Bradley}.   The less intense 0-1 sub-band is about 10\% broader and centered at 2.659($\pm 0.002$) eV for an offset energy of approximately 158 meV.  Because there are a multitude of Raman modes in the vicinity of 1300 cm$^{-1}$ a unique assignment is not possible (although there is a particularly strong band near 1280 cm$^{-1}$).  More likely it represents a superposition of modes with the strongest Huang-Rhys coupling parameters.  To represent these ``two'' sub-bands we follow the nomenclature of Gierschner et al.\cite{OPV:subband:jcp02Gierschner} and specify the two peaks, with approximate offsets of 160 meV and 200 meV, as $a_1$ and $b_1$ modes respectively.

Assignment of the three sub-band features in the 0-2 band is now straightforward.  These correspond to the three possible linear combinations of the $a_1$ and $b_1$ modes, $a_1$+$a_1$, $b_1$+$b_1$ and the mixed mode $a_1$+$b_1$ or, equivalently, $a_2$, $b_2$ and $a_1$,$b_1$ and absolute energies of approximately 2.50 eV, 2.42 eV, 2.46 eV.  The most intense of these belongs to the mixed mode band and this attribute is also seen in the recent OPV study\cite{OPV:subband:jcp02Gierschner}.  The 0-3 sub-band structure is only weakly resolved and this band consists of, at a minimum, four major sub-bands (i.e., $a_3$, $a_2$+$a_1$, $a_1$+$b_2$ and $b_3$).  Quantitative assessment of the F-C vibronic band lineshapes and positions clearly works best if there is prior knowledge of the underlying sub-band structure.  A simple analysis and comments on the thermal evolution of this sub-band structure will be presented later.

\subsection{Structure and Spectroscopy of the Order-Disorder Transition}

The discussion now shifts to the nature of the 80~$^\circ$C ODT and its relationship to the $\beta$ phase.  Structural phase transitions in polymers can be very sluggish and the transformation from the $\alpha$ phase to the $\beta$ phase on cooling is no exception.  In this case a spin-coated sample was first heated to 115~$^\circ$C while maintaining the sample in a N$_2$/toluene vapor environment and then moderately cooled (ca.\ 6~$^\circ$C/min) to a final temperature of 40$\pm 1~^\circ$C.  Thereafter the time evolution of both the PL and Abs were alternately recorded. Figure~3 contains a small subset of the sample C results.  This specific data is taken from light incident on the thinned (see Sec.~II) narrow strip between between the polymer free region and that of the full thickness spin-coated film.  We also observe a dependence of the ODT kinetics with varying PF8 polymer film thickness.  This will not be discussed further except to note that thinner films appear to have {\em faster} kinetics in contradiction to the behavior seen in thin-film $\sigma$-conjugated polysilanes\cite{pdhs:rabolt:96science,pdhs:rabolt:96mm}.  

\begin{figure}
\includegraphics[width=3.375in]{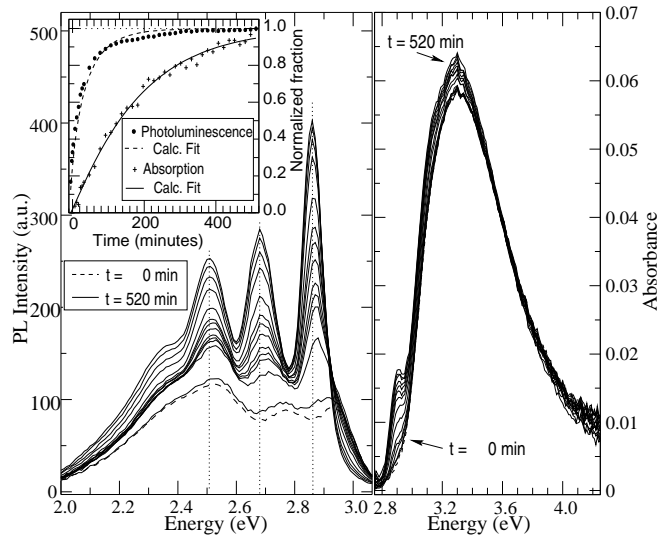}%
\caption{Selected absorption (right) and photoluminescence (left, $\lambda_{\mbox{\tiny{Ex}}}=400$~nm) spectra from PF8 sample C with respect to changes in time (at 40~$^\circ$C) after cooling from the thermotropic mesophase (to 115~$^\circ$C)  Inset: Relative fraction of the $\beta$ phase emission and absorption versus time (see text) in comparison with fits to Avrami type expression, $I(t)=1-\exp (-bt^n)$.\label{fig3}}
\end{figure}
 
Initially the PL consists mainly of very weak $\alpha$ phase 0-0 and 0-1 F-C bands superimposed on the broad background assumed to be from a variety of possible causes.  New, tell-tale emission indicative of the $\beta$ phase is resolved shortly after ($\approx 2$ min) reaching the thermal set-point.  PL from the 0-0, 0-1, 0-2 and the 0-3 F-C bands increases smoothly with time while the original $\alpha$ phase emission gradually decreases.  The underlying broad-band (BB) background emission also appears to increase slightly. The position of all three major $\beta$ phase vibronic bands red shifts $\approx 25$ meV with time.  The majority of this change occurs at early times.  This effect is consistent with a general increase in the backbone planarity and, concomitantly, enhancement in the effective conjugation length.  

In the case of absorption the changes are far less dramatic and only the low energy $\beta$ phase 0-0 band is measurably resolved.  All together these data are consistent with a simple two-phase coexistence framework.  There are other notable effects.  Not only does PL emission from the $\beta$ phase precede its appearance in the absorption data but the PL intensity increases at a more rapid pace.  This result is comparable to the progression seen in  solution studies of MEH-PPV\cite{MEH:twophase:mm01Rothberg} [(poly(2-methoxy,5-(2$'$ethyl)-hexyloxy-$p$-phenylene vinylene)] in which the solvent quality was step-wise varied to induce both aggregation and PPV backbone planarization.  Because the excitation wavelength is fixed to 3.11 eV (400 nm) absorption occurs throughout all $\beta$ phase portions and, to a lesser extent, in $\alpha$ phase regions as well.  A probable explanation for the anomalous PL intensity increase is that it is simply due to efficient energy migration of excitons along the polymer backbone and interchain energy transfer from regions having $\alpha$ type conformations to regions of $\beta$ phase.  

For a more quantitative analysis the following ad hoc procedure was implemented:  The relative proportion of $\beta$ phase PL with time, $f_{\beta,PL}(t$), was approximated by $$ f_{\beta,PL}(t) \equiv \frac{I_{PL}(t,E)-f_\alpha(t) I_{PL}(t=0,E)} 
{I_{PL}(t=500 \mbox{ min},E)-f_\alpha(t) I_{PL}(t=0,E)}$$ 
where $I_{PL}$ is the PL intensity at an energy $E$ which corresponds to the time-dependent intensity maximum of the $\beta$-phase 0-0 peak and $f_\alpha(t)$ is the relative remaining fraction of $\alpha$ phase.  Since PL emission at 3.00 eV is dominated by the $\alpha$ phase, this fraction is roughly given by  $f_\alpha(t) \equiv I_{PL}(t,3.00 \mbox{~eV})/I_{PL}(t=0 \mbox{ min},3.00 \mbox{~eV})$~.
The fraction of $\beta$ phase absorption, $f_{\beta,Abs}(t)$, was specified as 
$$f_{\beta,Abs}(t) \equiv \vspace*{.1in} \frac{I_{Abs}(t,2.90 \mbox{~eV})-I_{Abs}(t_0,2.90 \mbox{~eV})}{I_{Abs}(t=500 \mbox{ min},2.90 \mbox{eV})-I_{Abs}(t_0,2.90 \mbox{eV})}$$  where $t_0$ is the time
$t=0$. This approximation does not account for the time-dependent energy shifts in the 0-0 interband transition and so its intrinsic accuracy is lower.  

A plot of these normalized fractional components is shown in the Fig.\ 3 inset in conjunction with a simple Avrami type fitting function, $ f(t)=1-\exp(-b~t^n)$. This simple expression is often used to assess nucleation and growth\cite{pdhs:rabolt:96mm,ODT:avrami39jcp,ODT:avrami40jcp}.  The best-fit coefficients $b,n$ are (0.055, 0.8) and (0.0035, 1.08) for the PL and Abs curves respectively.  An exponent of $n=1$ is consistent with athermal nucleation followed by one dimensional growth.  In this case the growth direction occurs along the chain axis.  The overall increase in the effective conjugation length with time (as inferred from the red shift in the F-C peak positions) would be qualitatively consistent with this scenario.  The value of $b$ is associated with specific details of the local environment.  In terms of the overall ODT this is still an ongoing process even after ten hours have elapsed. 

The fit to the absorption data is clearly more consistent with the Avrami expression and this likely reflects the fact that all PL spectra include an additional emission contribution from excitons that have migrated from $\alpha$ phase regions. The overt PL curve shape change seen in the vicinity of 100 minutes may simply indicate the onset of competition between adjoining
$\beta$ phase domains for excitons which originate in $\alpha$ phase regions.  Any increase in the relative proportion of $\beta$ phase thereafter serves only to reduce the effective exciton migration path in the $\alpha$ phase necessary for reaching $\beta$ type regions.  A less likely explanation for this pronounced PL onset could be that $\beta$ phase regions which achieve the best structural order actually form first.  These sites would then function as the deepest system traps and thus the most PL efficient sites would actually form at early times in the ODT.  

By altering the energy of the excitation line it is possible to use PL spectroscopy as a spatially sensitive probe of the ODT.  The limited spectrometer resolution is problematic but useful results are still possible.  Figure 4 contains optical data using an identical thermal history as that in the preceding paragraphs except, in this case, the excitation line is now 2.91 eV (427 nm).  This guarantees that virtually all absorption is restricted to regions having transformed to the $\beta$ phase.  Thus, at $t=0$, there is little or no PL signal except for scattering of stray light from the excitation line.  Once again the PL emission precedes that of $\beta$ phase absorption.  Energy migration will take place within the $\beta$ phase and so this result is not unexpected. In these spectra the PL curves include only the three most intense F-C bands, 0-0, 0-1, and 0-2, the broad background and a much weaker 0-3 band emission as well. Interestingly the data suggests a stronger time-dependent red shift in the 0-0 band in comparison to that of the 0-1 and 0-2 bands.  The reason for this behavior is unknown.

\begin{figure}
\includegraphics[width=3.375in]{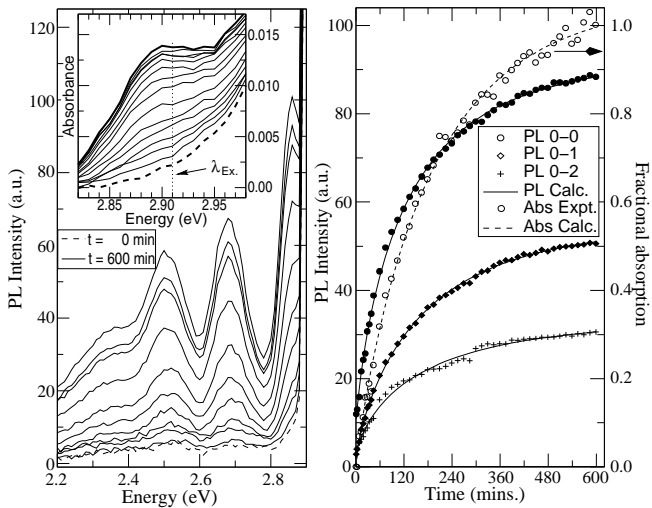}%
\caption{Selected photoluminescence (left, $\lambda_{\mbox{\tiny{Ex}}}=427$ nm) and absorption (inset, leading edge only) spectra from PF8 sample C (thin spot) with time (at 40~$^\circ$C) after cooling from the thermotropic mesophase (to 115~$^\circ$C).  At right: Relative fraction of the $\beta$ phase F-C vibronic band emission and absorption versus time (see text) in comparison with fits to Avrami type expression, $I(t)=1-\exp (-bt^n)$.\label{fig4}}
\end{figure}
 
One clear advantage is that each mode can be individually fit using a Gaussian profile and, subsequently, this intensity can be plotted as a function of time. Fitting of the 0-0 peak was not expected to be accurate because of the obvious overlap with the excitation line but good results were obtained nevertheless.  Now that all optical absorption has been restricted solely to regions of $\beta$ phase all three PL band intensities (and Abs curve as well) track the Avrami expression extremely well.  The coefficients are listed in Table I and these results are similar to those obtained previously. Since this ODT proceeds relatively slowly and smoothly we suggest that higher resolution time-resolved or SSF studies can yield further insight into this general process of energy migration within ``heterogeneous'' media.

\begin{table}%[H] add [H] placement to break table across pages
 \caption{Coefficients from Avrami type fit to expression 
$I(t)=1-\exp(-bt^n)$ for sample C as detailed in Fig.~3 (\#1) and 
Fig.~4 (\#2) \label{tab1}}
\begin{ruledtabular}
\begin{tabular}{lll}
Trial & ~~~~$b$ &  ~~$n$ \\  \hline
\# 1 Abs & 0.0035 & 1.08~~ \\
~~ PL ($\lambda _{\mbox{\tiny{Ex}}}=400$ nm)  & 0.055 &  0.80   \\ \hline
\# 2 Abs & 0.0057 & 0.99 \\
~~ PL 0-0($\lambda _{\mbox{\tiny{Ex}}}=427$ nm) & 0.031  & 0.72 \\
~~ PL 0-1 & 0.019  & 0.77 \\
~~ PL 0-2 & 0.030 & 0.68 \\ 
\end{tabular}
\end{ruledtabular}
\end{table}

Although the ODT kinetics are slow at 40~$^\circ$C they become progressively more rapid as the level of undercooling increases. This is a general phenomenon in polymers\cite{meta:98:keller,meta:98:cheng} and it continues until a glass transition is approached at which point the kinetics once again become more sluggish. This property has major implications for processing of PF8 thin films.  Although thermal cycling has been often used to improve the conversion to the $\beta$ phase one may expect that only a narrow temperature range is actually important for achieving this effect.  

To adequately address this issue it was first necessary to identify the lowest temperature range in which the PF8 ODT can proceed.  Some difficulties were initially encountered but the CO$_2$ quenching stage appears to have  effectively {\em suppressed} formation of the $\beta$ phase. Figure 5 displays a series of PL and Abs spectra on {\em warming} (ca.\ 2~$^\circ$C/min) of a previously quenched/anneal spin-cast film (sample C) after first heating to 390 K (before the quench) in a N$_2$/toluene vapor atmosphere.  The 110 K PL spectra contains little or no traces of the $\beta$ phase.  At this temperature the frozen $\alpha$ phase sample clearly exhibits a pronounced vibronic F-C progression with 0-0, 0-1 and 0-2 modes centered at the indicated positions.  The 0-1 and 0-2 sub-band structure is not resolved but its presence is inferred from the uneven energy spacing of the 0-1 and 0-2 phonon bands.  At temperatures below 240 K the overall PL lineshape and intensity variations resemble those of the branched PF2/6 polymer. The loss of  PL signal from the frozen $\alpha$ phase on warming is quite striking.

 \begin{figure}
 \includegraphics[width=3.375in]{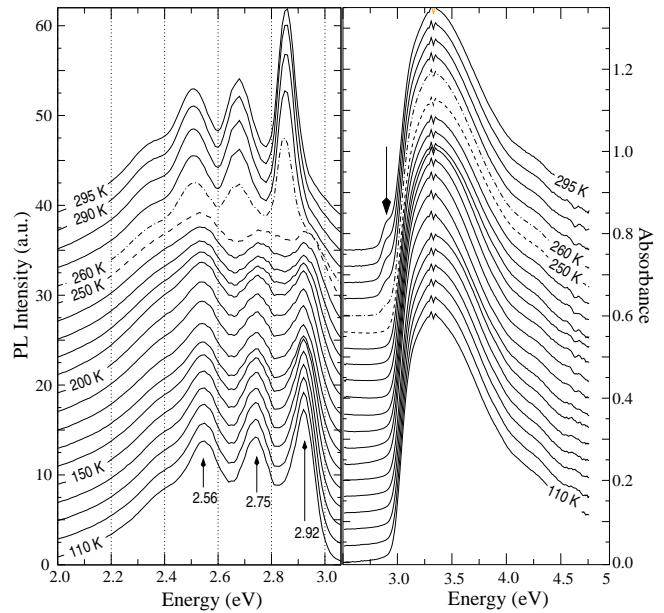}%
 \caption{Photoluminescence (left, $\lambda_{\mbox{{\tiny Ex}}}=400$ nm) and absorption (right) spectra from PF8 sample C (at a thick spot) on heating after quenching from thermotropic mesophase ($\alpha$ phase) and then cooling to 110 K.  The bold arrow identifies the 0-0 $\beta$ phase absorption band. All curves have been offset for clarity and the PL data has been corrected for self-absorption.  The anomalous absorption feature at 3.3 eV is an artifact due to changes in the light source intensity.\label{fig5}}
 \end{figure}

Dramatic change is observed in the PL emission after passing through 250 K.  By 260~K there has been a sharp increase in the PL intensity from chains now in the $\beta$ phase.  A well-defined shoulder is seen at 3.00 eV and this indicates the presence of limited residual emission from the remaining $\alpha$ phase regions.  Further study of the kinetics is necessary to better characterize this transition. The disparity in PL output between the two phases qualitatively suggests that the PL yield from the $\beta$ phase is significantly higher.  At this temperature there is also a slight hint of $\beta$ phase absorption, as indicated by the bold arrow, but the actual proportion of the $\beta$ phase regions is quite low.  Additional warming has no major effects except that the fraction of the $\beta$ phase  continues to increase somewhat.  The largest $\beta$ phase 0-0 emission intensity actually occurs in the 270 K spectrum.  Longer annealing times would be necessary to further reduce the relative proportion of $\alpha$ type emission.  Overall these data indicate that temperatures below 250 K are of no importance for conversion to the $\beta$ phase.

So far no data or discussion has addressed the relationship of crystallization and changes in interchain packing to the PF8 ODT.  The only relevant facts contained in the Fig.~5 data are that the actual proportion of $\beta$ phase is small and that the molecular level conversion to the $\beta$ phase occurs without significant changes in the broad background emission.  Figure 6 contains a series of XRD profiles from a PF8 film first on warming, after the initial casting, followed by slow cooling and then on warming after a more rapid cooling (but {\em not} a CO$_2$ spray quench).  The overall statistics are poor but the main intention in this study was to keep the film as thin as possible for comparison purposes to the optical spectra.  The as-cast film, initially cooled to -35~$^\circ$C, exhibits relatively poor structural order with only a single  sharp peak at $2\theta$=6.82$^\circ$ and a broader feature centered near 20.4$^\circ$.  These features are not consistent with earlier fiber data\cite{PFO:processXRD:mm99Bradley}.  They do however match results from a recent grazing-incidence (GI) thin film study\cite{PFO:tfXRD:polyBradley02} and, on the basis of limited information, the GI paper has proposed a triangular packing of the PF8 chains.  By 40~$^\circ$C there is weak indication of peaks at 13.1$^\circ$ and 15.3$^\circ$ but absolutely no peak intensity is detectable at either $\sqrt{3}$ or 2 times the wavevector of the fundamental $2\theta$=6.82$^\circ$ peak.  This strongly suggests that the PF8 polymer does not adopt a hexagonal columnar type packing in this ordered phase. 

 \begin{figure}
 \includegraphics[width=3.375in]{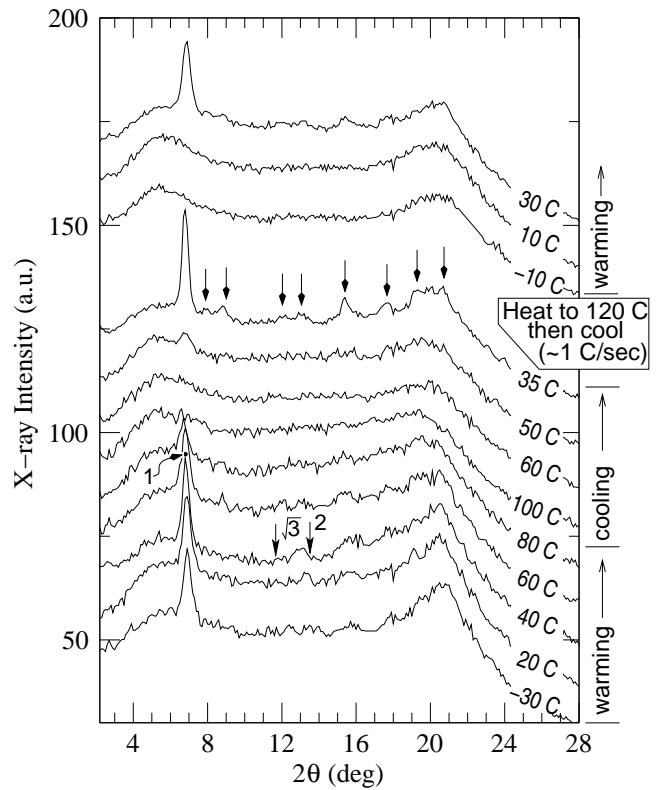}%
 \caption{X-ray powder diffraction profiles from a thick
film (cast from 0.5\% PF8 w/w THF/toluene mixture onto mica,  $\approx 5\mu$m thick) on warming and cooling as indicated.\label{fig6}}
 \end{figure}

Continued heating through the known thermotropic 80~$^\circ$C  transition to a PLC phase is paralleled by a complete loss of all sharp scattering features.  There is noticeable hysteresis and the crystalline phase  does not reappear on cooling until 50~$^\circ$C.  This attests to the very slow ordering kinetics for forming crystalline phases.  A long-time scan (8 hrs.) at 35~$^\circ$C reveals a series of Bragg peaks and these are listed in  Table II.  Indexing of these peaks proved difficult and, without knowing exactly which are equatorial and non-equatorial, we cannot tentatively assign an appropriate unit cell.  The large, 13 \AA~$d$-spacing likely corresponds to packing of the PF8 chains and the broad 4.3 \AA~feature arises from scattering by the alkyl side chains.  Once again a triangular lattice appears to be inconsistent with the data.

 \begin{table}%[H] add [H] placement to break table across pages
 \caption{Ordered phase $2\theta$ angles and $d$-spacings as indicated on the 35~$^\circ$C diffraction profile in Fig.~6.\label{tab2}}
 \begin{ruledtabular}
\begin{tabular}{rrrr}
$2\theta$~~ & $d$-spacing (\AA) &  $2\theta$~~ & $d$-spacing (\AA)\\  \hline
6.78  & 13.04~~~~~~ &	15.3 & 5.78~~~~~~ \\
7.87  & 11.24~~~~~~ &	17.6 & 5.03~~~~~~ \\
8.85  &  9.99~~~~~~ &	19.2  & 4.62~~~~~~ \\
12.04 &  7.35~~~~~~ &	20.7  & 4.29~~~~~~ \\
13.1~~ & 6.76~~~~~~ & \\
\end{tabular} 
 \end{ruledtabular}
 \end{table}

On even modest cooling from the PLC phase crystallization can be fully suppressed. Both the -10~$^\circ$C and 10~$^\circ$C diffraction profiles exhibit no traces of crystallization.  Only after warming to 30~$^\circ$C does the ordered phase return.  This temperature is substantially 
less than the 75~$^\circ$C reported PF8 glass transition temperature but any residual solvent in this film could function as a plasticizer and thereby lower this value.  We suggest that the reported glass transition corresponds only to temperatures at which intermolecular translational motions become possible. Intramolecular conversion to the $\beta$ phase occurs even at markedly lower temperatures.  Samples annealed at temperatures at or below the nominal 10~$^\circ$C threshold can develop a significant fraction of $\beta$ phase absorption.  We therefore conclude that conversion to the $\beta$-phase is a local, single chain relaxational process that will typically precede crystallization in the PF8 polymer.  For PF8 films the
presence of the $\beta$ phase conformation is a likely prerequisite for 
initiating crystallization.  This underlying single chain response is a  likely cause for reported improvements in light-emitting diode (LED) performance when annealing other polymers at temperatures below their respective glass transition temperatures\cite{LED:anneal:apl02cao}.  A high degree of interchain order is not essential for enhanced polymer emission properties.

\subsection{Temperature-Dependent Franck-Condon Vibronic Structure}

We now return to the discussion of the main F-C vibronic bands, the two $a_n$ and $b_n$ sub-bands and the overall temperature evolution.  Because two distinct vibrational sub-bands are resolved, all 0-$n$ peaks are fit assuming a constrained superposition of these two sub-bands, $a_n$ and $b_n$ (using Gaussian lineshapes), and all relevant cross-terms.  The most ideal case shown is that of sample C, as seen in Fig.~7 with the stated thermal history, and this sample exhibits very little (if any) $\alpha$ phase emission.  At 110 K the net emission at 2.92 eV is less than 2\% that of the $\beta$ phase 0-0 maximum.  In all other respects it matches the sample A PL in Fig.~2 except for a minimal red shift.  The $a_1$, $b_1$ sub-band contributions to the 0-1 lineshape remain distinct even at 110 K and higher temperatures. For curve fitting the intensity ratio of the three requisite 0-2 components approximates 1:2:1.  The relative intensities do change with temperature and the two unmixed sub-bands (i.e., $a_2$ and $b_2$) parallel the intensity variations in the 0-1 sub-bands.  For the four 0-3 sub-bands a nominal ratio of 1:3:3:1 was seen\cite{tracking}.  In all cases a smooth, slowly varying background curve was necessary.  This is labelled ``BB'' in the figure and it originates from a variety of possible effects (as mentioned earlier in the text).  Finally we note that it was also necessary to include a weak but localized peak centered about 2.76 eV.  This feature has been identified previously as a vibronic replica\cite{mLP:emigration:cplett00List,PFO:PLmany:prb02Guha}.% 

 \begin{figure}
 \includegraphics[width=3.375in]{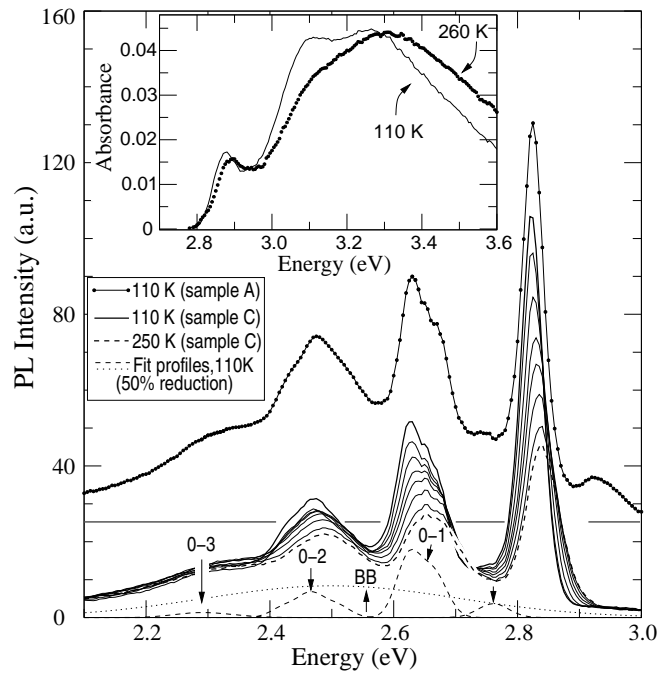}%
 \caption{Sample C photoluminescence (from a thin spot) on  warming from 110 K to 250 K in comparison with PL data of sample A (at 110 K) and, in addition, the various indicated modeling profiles from fit to 110 K spectra (dashed lines after a 50\% reduction). In this instance sample C was first heated to 380 K (in a N$_2$/toluene atmosphere), quenched to -40~$^\circ$C, annealed ca.~1 hr at 10~$^\circ$C and then annealed ca.~1 hr at 40~$^\circ$C.  These PL data also include self-absorption corrections. The sharp feature at 2.650 eV is a detector artifact.\label{fig7}}
 \end{figure}

The major PL response to heating are the systematic changes in all peak widths and positions combined with a gradual loss of the F-C emission.  In terms of the sub-band structure only variations in the relative 0-1 sub-band intensities are of real quantitative significance and, surprisingly, these are strongly temperature dependent.  At the lowest temperature the $b_1:a_1$ intensity ratio is about 3:2 whereas, at 260 K, the curve fitting required a near 1:1 ratio.   This progression may imply subtle underlying changes in the Huang-Rhys coupling to the various vibronic states or, more likely, the effects of cross-coupling and anharmonicity.  
Recent Raman scattering studies\cite{ltraman} of the PF2/6 derivative report a systematic increase in the 1600 cm$^{-1}$ band intensity on cooling in qualitative agreement with these PF8 PL results.  

Changes in temperature also strongly affect both the 0-0 PL peak line shape and position.  At lower temperatures emission occurs increasingly at chain segments having the smallest interband $\pi-\pi^*$ transition energies. This causes both a strong red-shift and narrowing of the 0-0 in PFO polymers\cite{PFO:F8morph:prb00Bradley}.  One qualitative and often cited explanation is that the freezing out of low energy librational and vibration modes increases the effective conjugation length.  A cogent counter argument\cite{SSF:review:acr99Bassler}, based on the observation that polymer emission strongly resembles that of the oligomers, discounts this mechanism and states that this behavior simply reflects the temperature-dependence of electronic relaxation processes.  The actual thermal evolution is sensitive to both the choice of polymer\cite{PFO:PLmany:prb02Guha} and the physiochemical processing history.  Figure 8 displays the peak position of the 0-0 band versus temperature for a variety of different PF8 samples and thermal preparation conditions.  Thermal history clearly has a very strong impact.  In the best cases we observe a large linear temperature dependence of 1.5$\times 10^{-4}$ eV/K.  In many instances however there are deviations from linearity with a far more complicated temperature dependence.  

 \begin{figure}
 \includegraphics[width=3.375in]{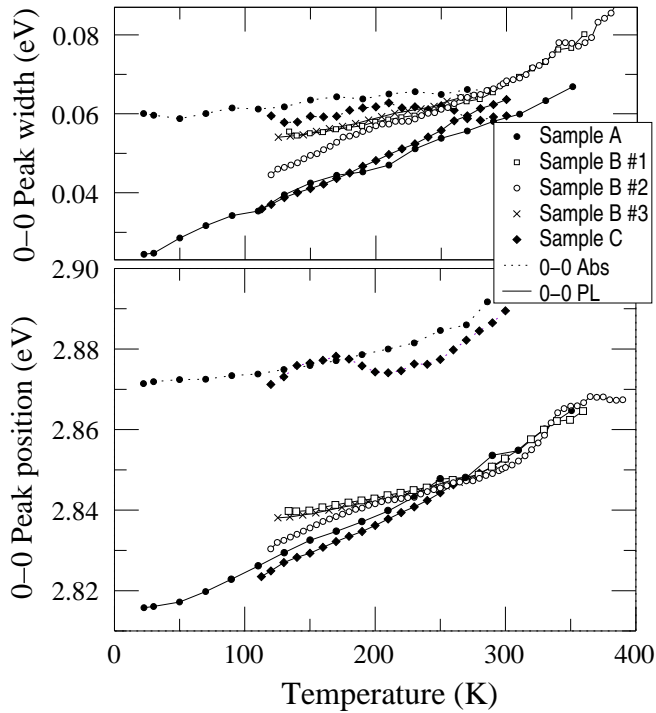}%
 \caption{$\beta$ phase 0-0 PL and Abs peak positions (bottom) and width (top, full-width at half-maximum) versus temperature for selected samples and thermal histories.  They are as follows: $\bullet \rightarrow$  sample A (see Fig.~2), filled-$\diamond$ $\rightarrow$  sample C (using conditions stated in the Fig.~7 caption), ${\mbox{ \tiny $\Box$}} \rightarrow $ sample B (\# 1, in vacuum, heat to 390 K, cool briefly to ca.~273 K, 3 days at 293 K), $\circ\rightarrow $   sample B (\# 2, ca.~24 hr after casting at 293 K),  $\times\rightarrow $  sample B (\# 3, in vacuum, heat to 390 K, five days at 70 K , 24 hr at 293 K).  With the exception of sample A all data is recorded during warming from low temperature.\label{fig8}}
 \end{figure}

Comparisons to the 0-0 peak width progression are even more striking because the temperature dependence of each sample strongly tracks that of its own peak position shift.  After accounting for the finite spectrometer resolution even the magnitudes of these changes have a near one-to-one correspondence.   Large net shifts are observed and, in the case of sample A, a 50$\pm2$ meV displacement occurs over the 325 K temperature range.  The evolution of the 0-0 absorption peak position is generally much weaker and its peak width remains nearly constant.  If net increases in the effective conjugation were dominant then one should expect a qualitatively similar response.  This strong temperature-dependent PL behavior is ostensibly more consistent with the electronic relaxation hypothesis.  

A strong correlation between PL width and position can be modeled by assuming that emission originates from excitons in thermodynamic equilibrium  in states at the bottom of an exciton band.  Thus emission is proportional to a product of the density of states weighted by the oscillator strength (or $\cal D (\varepsilon)$) times a Boltzmann factor to give ${\cal D} (\varepsilon) \exp (-{\varepsilon} /k_B  T) $ in close analogy to emission from organic molecular crystals\cite{MXTAL:exciton:jcp68colson,MXTAL:exciton:jcp70hanson}.  In this  case we expect that intrachain, not interchain, interactions should dominate the exciton band.  An expression with this form will produce noticeable asymmetry in the 0-0 lineshape and this attribute can be observed in our experimental data\cite{satisfactory}.  A correlated linear dependence of the peak widths and positions requires a power law density of states and, in this case, we find that an expression  $ A(T)~T^{-5/2} \varepsilon^{3/2}~\exp(-\varepsilon/k_B T) $ qualitatively reflects the experimental results where $A(T)$ incorporates the temperature dependence of the integrated intensity and the factor of $T^{-5/2}$ provides normalization.  With an exponent of $\frac{3}{2}$ the shift in peak position is equal to $\frac{3}{2}k_B T$ and the full-width at half-maximum is $3k_B T$ and this result is in approximate agreement with the data shown in Fig.~8. Extrapolation to zero width gives a threshold energy, $\varepsilon_T = 2.800\pm 0.003$ eV, so the full expression, ignoring the effects of instrumental resolution and polymer inhomogeneities, becomes
$$ PL(\varepsilon, T) = A(T)~T^{-5/2}~(\varepsilon-\varepsilon_T)^{3/2} \exp (-(\varepsilon-\varepsilon_T)/k_B T)$$
for when $ \varepsilon \ge \varepsilon_T.$  Comprehensive analysis would require a composite function incorporating a distribution of chain conformations to reflect residual disorder and the presence of additional low energy F-C vibrational modes\cite{PPV:trSSF:prb01phillips,PPV:inelastic:prb95} (on the low energy side).  A threshold value of 2.80 eV is still more than 0.13 V higher than the 0-0 peak position in the planarized ladder-type poly($p$-phenylene) polymers\cite{mLP:emigration:cplett00List} and therefore we conclude that even in the $\beta$ phase the PF8 backbone adopts, on average, a non-planar conformation.

The existence of a 2.80 eV threshold from the $\beta$ phase PL is also consistent with the absorption results when they are extrapolated to zero temperature.  In this case we observe that the  $\beta$ phase 0-0 absorption peaks at 2.87 eV and has a full-width at half-maximum of 0.06 eV. The onset
of this interband absorption occurs at ~(2.870-0.060) eV or 2.81 eV which
is just above the emission energy threshold.  All together these results and
the modeling  reinforce the claim that much of the bathochromic shift originates from electronic relaxation processes.

Better resolution of the underlying sub-band structure facilitates more general interpretations of the 0-1 and higher order vibronic peaks.  For example, if the relative intensity variations between the two resolved sub-bands are not taken into account then one would obtain temperature-dependent changes of the 0-1 position which are anomalous.  This effect may be, in part, responsible for the differing 0-0 and 0-1 slopes as reported in Ref.~\onlinecite{PFO:PLmany:prb02Guha}.  In our samples we observe only very, small temperature-dependent shifts of the 0-1 $b_1$ sub-band energy (at top in Fig.~9) with a coefficient of no more than $\approx 1\times 10^{-5}$ eV/K.  This result is {\em very} sensitive to the exact form of the fitting function.  An increase in the sub-band energy with reduced temperature is consistent with an increase in ``stiffness'' of the surrounding media.  {\it In situ} Raman measurements generally show a gradual increase in the mode energies with reduced temperature\cite{PFO:F6F8raman:jap02bradley} as well.   

 \begin{figure}
 \includegraphics[width=3.375in]{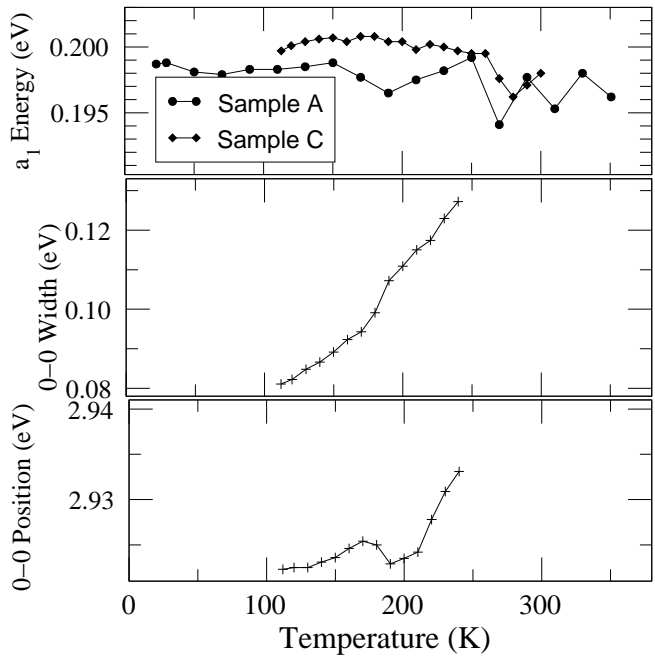}%
 \caption{Above: Energy of $\beta$ phase $b_1$ sub-band versus temperature for samples A and C (see Figs.~2 and 7 for conditions). Below: Frozen $\alpha$ phase (i.e., sample C after quench) 0-0 peak width (full-width at half-maximum) and position as a function of temperature.\label{fig9}}
 \end{figure}

A slightly more rigorous analysis of the relative peak widths is also possible.  Figure 10 shows the evolution of the underlying 0-$n$ sub-band peak widths versus temperature from the two trials that obtained a uniformly linear response. All peaks widths are comparable and generally exhibit the same linear trend on warming.  At temperatures below 200~K, where the individual sub-band peaks can distinguished, the intrinsic 0-0, 0-1 and 0-2 component peak widths are typically within 20\% of one another.  Anomalous behavior (i.e., in which the 0-1 and 0-2 widths are less than that of the 0-0 peak) are attributed to artifacts in the curve analysis due to uncertainties in the broad background line shape\cite{better} and the non-Gaussian profile of the 0-0 emission band.  Alternatively, one can rigidly specify the 0-0 peak position, a net Huang-Rhys (H-R) coupling parameter, a single sub-band peak width and  the relative H-R sub-band contributions for the two resolved sub-bands (plus their positions) for a total of only 6 free parameters.  With only minor modifications to account for systematic effects (e.g., the $\approx 10$\% width difference between the $a_1$ and $b_1$ sub-bands) and a slowly varying background profile we obtain almost equally good fits to the experimental data (not shown).  For sample C, at 110 K, an H-R parameter of $0.63\pm 0.03$ is obtained and it increases gradually to approximately 0.8 at temperatures near 300 K.  This progression is similar to that reported for the PF2/6 polymer\cite{PFO:PLmany:prb02Guha} although the PF2/6 polymer conformation is more $\alpha$ phase like.

 \begin{figure}
 \includegraphics[width=3.375in]{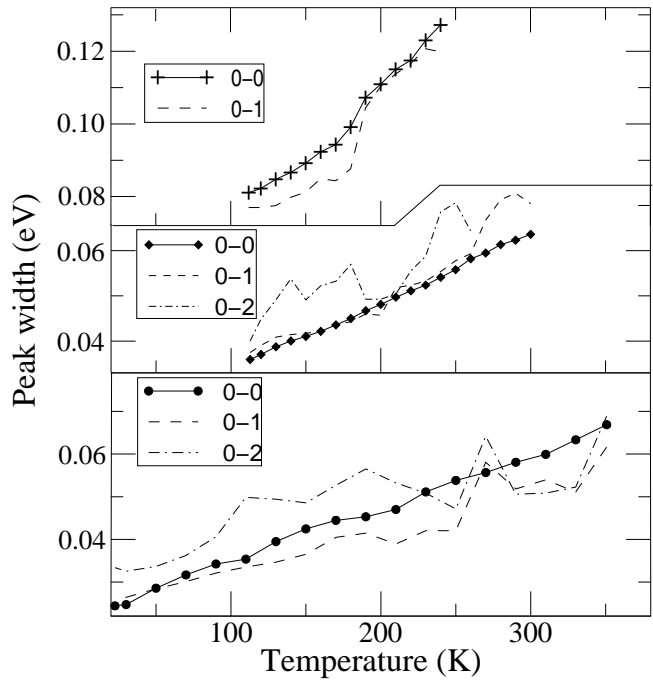}%
 \caption{Underlying vibronic 0-$n$ sub-band widths (full-width at half-maximum) versus temperature for sample A $\beta$ phase (bottom, see Fig.~2 for conditions), sample C in the $\beta$ phase  (middle, see Fig.~7 for conditions) and quenched sample C in the $\alpha$ phase (top).\label{fig10}}
 \end{figure}

The PL temperature dependent behavior of the frozen PF8 $\alpha$ phase film (i.e., sample C after quenching) can be analyzed as well using the aforementioned treatment and these results are shown in both Figs.~9 and 10.  In this case the 0-0 peak position shift is not linear over the accessible temperature range and the net width change is more than twice as great as that of the energy shift.  At temperatures near 250 K, the point at which rapid conversion to the $\beta$ phase initiates, there is increased uncertainty in the curve fitting because there is some indication that there may be more that one spectral feature comprising the 0-0 PL peak.   Including the sub-band structure in the curve fitting gives, at 110 K , a $b_1$ offset energy of 205($\pm 5$) meV.   This value is somewhat higher than that of the $\beta$ phase.  Raman studies\cite{PFO:Ramanmorph:sm00Bradley} have reported a small drop from 1605 cm$^{-1}$ to 1602 cm$^{-1}$ on PF8 crystallization; a result qualitatively consistent with this observation.  Naively one might expect crystallization to increase the elastic constants.  The 0-1 PL peak width progression is relatively linear and its magnitude is $\approx 3 \times 10^{-4}$ eV/K; a value approximately twice that of the $\beta$ phase.  The H-R parameter is close to 0.75 at 110 K and this parameter also increases somewhat with temperature. Without narrower 0-$n$ lineshapes, better knowledge of the broad background and more clearly resolved F-C modes in the absorption spectra further interpretation is unwarranted.

\section{Conclusions}

The presence of mesomorphic behavior in conjugated polymers provides a unique opportunity for probing structure/property relationships and their influence on both charge transport and photophysics.  This claim must be strongly tempered by the fact that both residual disorder and the intrinsic nature of this phase behavior can generate a diverse set of results.  In the PF8 polymer it is the sluggish nature of the ODT and the persistence of processing related heterogeneities which creates significant obstacles for experimental studies. 

Through a combination of temperature and time dependent studies we have
documented many of these effects and have gained additional insight into the nature of the ODT in PF8.  By approaching the ODT on both cooling and  heating it is possible to fully isolate the effects of local intrachain relaxation at the molecular level from those that pertain to interchain motion and subsequent crystallization.  Quenching experiments unambiguously show that formation of the $\beta$ phase corresponds first and foremost to intrachain relaxation. Once the polyfluorene chains have adopted this more planar conformation then interchain ordering and crystallization can progress.  Similar behavior is seen in alkyl-substituted polysilanes\cite{Psil:withoon:mm00xrd}.  On heating through the ODT the loss of intrachain and interchain structural ordering is well correlated.

Samples with improved structure order at the molecular level have enabled a more complete description of the F-C vibronic progression in the emission spectra.  Higher order 0-$n$ bands include strong contributions by the mixed-mode cross terms and this behavior parallels that of recent OPV studies. Analysis of the 0-0 PL peak shows features indicative of emission from excitons which are thermally equilibrated.  This result is very strong evidence in support of prior claims that electronic relaxation processes are the dominant mechanism for bathochromic shifts in conjugated polymer PL.  However the evidence presented in this work is identified with an electronic band and, if this behavior truly arises from intrachain interactions, therefore highlights a polymeric attribute.  Analysis of the 0-0 data could be further extended in future studies to directly extract the quasi-one-dimensional exciton band dispersion for comparison with {\it ab initio} quantum chemical methods\cite{PPV:conform:prl02saxena} which can derive excited state geometries.

\begin{acknowledgments}
We gratefully acknowledge NSF support of this work through grant DMR-0077698. One of the authors (M.J.W.) has benefited greatly from illuminating and fruitful discussions with Egbert Zojer, Suchi Guha, Niels Harrit and Donal Bradley.  Egbert Zojer and Avadh Saxena are both acknowledged for critical readings of the manuscript.  We thank Suchi Guha and coworkers for communication of their work prior to publication.  We also acknowledge Withoon Chunwachirasiri for assistance in acquiring the spectroscopic data. 
\end{acknowledgments}

\bibliography{/home/winokur/bib/lcp,/home/winokur/bib/kin,/home/winokur/bib/ppv,/home/winokur/bib/oops,/home/winokur/bib/psil,/home/winokur/bib/pfo}

\begin{thebibliography}{66}
\bibitem[{con()}]{conformer}
\bibinfo{note}{Throughout this paper we use $\alpha$ and $\beta$ to represent
  the two known PF8 backbone conformations and do not associate these
  references to a specific crystalline phase.}
\bibitem[{les()}]{lessdisorder}
\bibinfo{note}{A small reduction in the molecular weight may actually help
  reduce molecular entanglement as a source of inhomogeneous broadening}.
\bibitem[{bet({\natexlab{b}})}]{better2}
\bibinfo{note}{This represents an upper limit. If there are Lorentzian
  components to the lineshape then the extrapolated width would be even lower}.
\bibitem[{sat()}]{satisfactory}
\bibinfo{note}{The assumption of a Gaussian line shape is a reasonable
  approximation but close inspection show clear systematic failings. Higher
  spectrometer resolution is necessary for a thorough and proper analysis.}
\end{thebibliography}

\end{document}